# Generalized quantum interference of correlated photon pairs

Heonoh Kim, Sang Min Lee, and Han Seb Moon*

Department of Physics, Pusan National University Geumjeong-Gu, Busan 609-735, Korea

*Email: hsmoon@pusan.ac.kr

## Abstract

Superposition and indistinguishablility between probability amplitudes have played an essential role in observing quantum interference effects of correlated photons. The Hong-Ou-Mandel interference and interferences of the path-entangled photon number state are of special interest in the field of quantum information technologies. However, a fully generalized two-photon quantum interferometric scheme accounting for the Hong-Ou-Mandel scheme and path-entangled photon number states has not yet been proposed. Here we report the experimental demonstrations of the generalized two-photon interferometry with both the interferometric properties of the Hong-Ou-Mandel effect and the fully unfolded version of the path-entangled photon number state using photon-pair sources, which are independently generated by spontaneous parametric down-conversion. Our experimental scheme explains two-photon interference fringes revealing single- and two-photon coherence properties in a single interferometer setup. Using the proposed interferometric measurement, it is possible to directly estimate the joint spectral intensity of a photon pair source.

Nonlocal quantum interferences between strongly correlated photons are of fundamental interest not only for obtaining a deeper understanding of quantum mechanics, but also to explore the field of quantum information science. The observation of nonlocal two-photon interference effects is now considered as one of basic steps toward implementing photonic quantum information technologies. The fundamental principle underlying modern quantum optical technologies utilizing correlated photon pairs is that a photon-pair interferes with the pair itself, when the indistinguishability of two-photon probability amplitudes is guaranteed in coincidence detection. Representative two-photon quantum interference phenomena include the Hong-Ou-Mandel (HOM) interference effect arising from various conditions[1-3] and interferences of the path-entangled photon-number state (N00N state)[4-6], which lie at the heart of entanglement-based quantum communication and super-resolution metrology.

Since the original HOM experiment was performed, various types of HOM experiments have been performed to explore the relation between coherence and indistinguishability[7-12]. From the conventional framework, the overlap of two photons at beamsplitter (BS) has been usually accepted as basic concepts to understand the HOM effect. The extended version of the HOM-type two-photon interference experiment has been performed by adopting the superposition state as an input[3]. It has been shown that the spatial two-photon bunching effect can be observed even when the incoming photons have no identical properties such as polarization and arrival time. The concept of 'non-overlapping' in observing two-photon interference effect could also be applied to demonstrate the quantum interference between N00N states. The N00N state is a two-mode path-entangled N-photon state where all identical photons spatially/temporally bunched up at the same mode. Recently, it was reported that the path-entangled N00N-like two-photon state (time-like separated two-photon N00N state), which is easily generated from Greenberger-Horne-Zeilinger states, can enhance metrological resolution under strict condition of large separation between constituent photons[13]. The main issues of the previous works were obviously focused on the 'overlap of two interfering photons' and 'bunching up in the same spatial/temporal modes'. Furthermore, the two-photon interference effects observed in the HOM schemes have been treated distinctly different from that of involving path-entangled states. Although there are many literatures reporting the investigation of two-photon quantum interference phenomena, a fully generalized two-photon quantum interference scheme accounting for the HOM scheme and path-entangled N00N-like two-photon states has not yet been proposed thus far.

In this paper, we present an experimental demonstration of the most generalized two-photon interference with photon pairs that are independently generated by spontaneous parametric down-conversion in the telecommunication wavelengths. The proposed two-photon interferometric scheme has the following properties. (a) The local behavior of individual single photons in a correlated pair at the BS is completely independent so that the measurement of photons has to be performed under strict local condition (nonlocal interference)[14]. (b) The two-photon amplitudes are kept separate to observe phase-sensitive and/or phase-insensitive interference fringes[15]. (c) Constituent photons consisting N00N-like two-photon state are not bunched up in the same spatial mode (space-like and time-like separation)[13]. (d) Single-photon and two-photon coherence properties are simultaneously observed in a single interferometer setup. In particular, our two-photon interferometric scheme can provide a new interferometric method to measure the joint spectral intensity of the two-photon source via the inverse Fourier cosine transform of the full interferogram[16-18].

## Results

**Schemes for two-photon interferometer.** To observe the nonlocal two-photon interference effect in an interferometer, it is required that the two interfering photons (two correlated photons) do not meet at the same BS simultaneously. Rather, the two photons have to be separated sufficiently far from each other to satisfy the Einstein locality condition, i.e., the travel time difference between the two photons must be longer than the coincidence time window[14]. However, the two-photon state has to be in a superposition state regardless. Figure 1a shows a typical HOM-type two-photon interferometer with a large time delay between two input photons, but the input photons are in a symmetrically superposed state in space and time[3]. The gray bars represent the correlated photons in a pair. As is well known, the HOM interference effect is interpreted as the destructive interference of two two-photon probability amplitudes corresponding to the coincidence detection events of transmitted-transmitted (TT) and reflected-reflected (RR) cases[1]. In general, the coincidence probability in the HOM experiment is described as

$$P_{\text{coin.}} = \frac{1}{2}\left\{1 - V\left[f(\delta t)\cos(\Delta\omega\delta t)\cos\theta\right]\right\}, \qquad (1)$$

where $V$ denotes the visibility of the coincidence fringe and $\delta t$ the time delay between two input ports of the BS and $\Delta\omega$ the angular frequency difference of two input photons. The filter function $f(\delta t)$ is related to the spectral shape of the detected photons. The oscillation $\cos(\Delta\omega\delta t)$ can be observed when the two incoming photons are in a frequency-entangled state[7,19], and $\cos\theta$ is a continuous modulation term if there is a relative phase difference $\theta$ between the TT and RR amplitudes. Almost all the experiments performed thus far have not recorded the phase-sensitive HOM interferogram resulting from Eq. (1), because in these experiments, the TT and RR amplitudes of the two-photon states were overlapped in the same interferometer setup[15]. When the BS is separated to control the two amplitudes individually, a phase-sensitive HOM interference fringe can be observed.

Figure 1b depicts an interferometer involving the N00N-like two-photon state (a fully unfolded version of the path-entangled N00N state) that leads to neither bunching of two photons in the same spatial mode nor overlapping at the same BS. Unlike in the case of the conventional path-entangled state (all identical photons are in the same spatial mode at a time), spatially separated two photons do not arrive at the same BS at the same time in this configuration. Nevertheless, indistinguishable two two-photon probability amplitudes leading to coincidence detection can cause two-photon interference. It has been known that an unbalanced arrival time between two photons is not the critical restriction to observe the two-photon interference effect[3,20,21]. Indeed, the separation time (or distance) between the two correlated photons does not affect the interference fringe[13]. If the BS in Fig. 1a is separated into two BSs (BS1 and BS2), then we can construct the most generalized two-photon interferometry including the HOM scheme and N00N-like two-photon state in a single interferometer setup. Furthermore, in this case, we cannot distinguish between configuration of the HOM scheme and the interferometer involving the N00N-like two-photon state. In the case that one of the four paths is controlled individually, the phase-sensitive HOM fringe in Eq. (1) can be obtained in coincidence detection with two detectors (for example, D1 and D3). On the other hand, if the two paths are adjusted simultaneously, then the two-photon interference fringe reveals two-photon coherence property with phase super-resolution[22]. Although the HOM is an explicit two-photon interference phenomenon that originates from interference between two two-photon probability amplitudes, an examination of the full fringe pattern reveals the single-photon coherence property irrespective of the phase sensitivity or phase insensitivity of the fringe pattern only when there is no frequency entanglement between two photons in a pair. On the other hand, in the case of N00N-like two-photon interference, the full interferogram exhibits pump beam coherence or the two-photon coherence property, whether or not the two constituent photons arrive at the same BS or not.

Now, we propose a conceptual schematic to explain the generalized nonlocal two-photon interference between separate photon pair sources, as shown in Fig. 1c. Let us consider only the case of single photon pair generation from a separate source $\Phi_A$ or $\Phi_B$. If these two emission processes are coherent, then the quantum state of the total system can be described as $|\Psi\rangle = \frac{1}{\sqrt{2}}(|\Phi_A\rangle + e^{i\phi}|\Phi_B\rangle)$, where $\Phi_A \rightarrow |\omega_1(S_A)\rangle|\omega_2(L_A)\rangle$, and $\Phi_B \rightarrow |\omega_1(S_B)\rangle|\omega_2(L_B)\rangle$. Here $\phi$ represents the relative phase between the two processes. $S_i$ and $L_i$ are short path and long path of the source $\Phi_i$. $\omega_1$ and $\omega_2$ are angular frequencies of the photons in $S_i$ and $L_i$ paths, respectively. The two correlated single photons ($\omega_1$ and $\omega_2$) from a pair do not meet at the same BS at the same time. Thus, the local behavior of the individual photons at each BS is completely independent. However, the simultaneous behavior of two photons in a pair is mutually correlated, because the two possible amplitudes leading to coincidence detection are superposed[22]. If we detect the coincident photons with two single-photon detectors D1 and D3, then the final state just before detection can be written as

$$|\Psi\rangle = \frac{1}{\sqrt{2}}\left(|\omega_1(S_A)\rangle_{D1}|\omega_2(L_A)\rangle_{D3} + e^{i\phi}|\omega_1(S_B)\rangle_{D1}|\omega_2(L_B)\rangle_{D3}\right). \qquad (2)$$

Since the phase factor ϕ depends on the relative path-length difference between the short and long paths, it can be adjusted by varying the path-length difference, $|L_A - L_B|$, when the short path is fixed ($S_A = S_B$). Eq. (2) does not contain any constraint on the emission/travel time difference between the two photons. It is worth noting from Eq. (2) that traveling time of one photon ($\omega_2$) in a pair can be considerably longer than that of the other photon ($\omega_1$) when compared with their coherence time (single- and two-photon coherence time). Moreover, this time can also be considerably longer than the coincidence time window[14].

**Experimental setup.** The experimental setup to demonstrate the proposed two-photon interference is shown in Fig. 2. Correlated photon pairs at 1.5 µm telecommunication wavelengths are generated through the quasi-phase-matched spontaneous parametric down-conversion (QPM-SPDC) process in type-0 periodically-poled lithium niobate (PPLN) crystals. We use a mode-locked picosecond fibre laser (PriTel, FFL-20-HP-PRR and SHG-AF-200) as a pumping source for QPM-SPDC, whose pulse duration is 3.5 ps at the center wavelength of 775 nm with a repetition rate of 20 MHz. Pump pulses amplified by an erbium-doped fibre amplifier (EDFA) are mode-locked in an intra-cavity and fed into a 1 mm-long PPLN crystal for frequency doubling. The second harmonic pulses are divided by the BS, and the outputs are used to pump the two 10 mm-long PPLN crystals. In our experiments, the average pump power was set as 20 mW for each crystal. Nondegenerate (1530 nm and 1570 nm) photon pairs are emitted with the full-opening angle of 4.6° in the noncollinear regime (see Methods for details). Broadband down-converted photons are coupled to a single-mode fibre (SMF) whose coupling bandwidth was measured as about 194 nm via HOM dip measurement without spectral filtering.

Our experimental scheme is composed of two fibre interferometers as depicted in Fig. 2. A nonlocal interference under the Einstein locality condition can be observed even when the total traveling time of the two photons is considerably longer than the coincidence resolving time (10 ns in our experiment). To satisfy this condition, an additional fibre delay line (6-m SMF) is inserted in the outer interferometer. The fibre length of the inner interferometer is about 4 m. Optical delay lines (ODL1 and ODL2) are inserted at each interferometer to adjust the path-length differences within single- and two-photon coherence lengths. With fine adjustment of ODL2, we observed phase-sensitive two-photon interference fringes, which are depicted in Fig. 3b.

Correlated photons generated from a PPLN crystal do not overlap at the same BS. Individual photons from the two sources are combined at two spatially separated fibre BSs (FBS1 and FBS2) to interfere nonlocally. To obtain the full interferogram of the single-photon and the two-photon wave packet, ODL 2 is scanned in the range of single-photon coherence length for various positions of ODL1 within the two-photon coherence length, as shown in Fig. 4a. Single photons are detected after they pass through coarse wavelength-division multiplexing (CWDM) filters via four InGaAs/InP single-photon avalanche photodiodes (APDs) operated in the gated mode. Electronic trigger signals are sent from the pump to the gate of the detectors by way of electric delay lines. The trigger signals are lowered from 20 MHz to 4 MHz to achieve the maximum external trigger rate of the four detectors. The quantum efficiency of detectors is set as 15%.

## Discussion

For alignment of the interferometer, the HOM two-photon interference between independent photons from the two PPLN crystals was observed in the two-fold coincidence measurements with detectors D1 (D3) and D2 (D4) as a function of the path-length difference $\Delta x_1$ ($\Delta x_2$) while varying the path-length difference with ODL1 (ODL2). Figure 3a shows the two-fold raw coincidence counts between D1 and D2. Each point represents the averaged raw data of two-fold coincidence counts per 10 s, and the error bars represent the counting uncertainties over 10 trials. Next, to observe nonlocal two-photon interference between the photon-pair amplitudes from two independent sources, it is required to match all of the four paths within their coherence length, as shown in Fig. 2. We measured the two-fold coincidences with detectors D1 and D3 by scanning $\Delta x_2$ (step size of 1 μm) under the condition of $\Delta x_1 = 0$. The coincidence detection probability according to the path-length difference $\Delta x_2$ has the form

$$P_{\text{coin.}} = \frac{1}{2}\left[1 - V\mathrm{sinc}\left(\frac{\Delta x_2}{\sigma_x}\right)\cos\left(\frac{2\pi}{\lambda}\Delta x_2\right)\right], \qquad (3)$$

where $V$ represents the fringe visibility and $\sigma_x$ the fringe width related to the single-photon spectral bandwidth and $\lambda$ the single-photon wavelength. Figure 3b shows the measured interference pattern, which is the most general HOM type two-photon interference fringe including the relative phase difference between two amplitudes in Eq. (2)[15,21]. Spatially separated photons of a pair exhibit nonlocal behavior with the given interferometer because the two-photon amplitudes are superposed (Fig. 1c). The phase-sensitive full fringe pattern ($\pi\sigma_x = 0.141$ mm) reveals the single-photon coherence property ($\lambda^2/\delta\lambda = 0.137$ mm). To show the phase sensitivity, we measured the fringe at $\Delta x_2 \approx 0$ (step size of 150 nm) as shown in Fig. 3c. The error bars represent the standard deviation assuming a Poissonian distribution for the counting statistics. Accidental coincidences are subtracted from the raw data, which is simply calculated from the equation $N_{\text{acc.}} = N_1 N_3 / f_{\text{trig.}}$, where $N_1$ and $N_3$ the single photon counting rates are 95 kHz and 91 kHz, respectively, and $f_{\text{trig.}}$ the trigger frequency for each detector is 4 MHz in our experiment. The gray area in Fig. 3b corresponds to the interference fringe obtainable by using rectangular-shaped spectral filters, corresponding to Eq. (3). High visibility of the two-photon interference fringe is observed even though the two photons in a pair do not meet at the same BS simultaneously. Furthermore, the difference between the travel times of two photons is longer than the coincidence time window.

Upon introducing an unbalanced path-length difference using $\Delta x_1$ and $\Delta x_2$ (Fig. 2), we can observe the full interferogram of the N00N-like two-photon state including single-photon and two-photon coherence properties. In this case, the fringe visibility is strongly dependent on the symmetry between the two amplitudes in Eq. (2), and the two-photon interference fringes are functions of the path-length differences $\Delta x_1$ and $\Delta x_2$, as shown in Fig. 4a. As expected, the which-path information originating from the geometric asymmetry degrades visibility within the range of the two-photon coherence length. Figure 4b shows the fringe visibilities as a function of $\Delta x_1$. Maximum visibility is observed when individual photons from different pairs are exactly matched at each BS within the length of the single-photon wave packet ($\Delta x_1 = 0$ and $\Delta x_2 = 0$). The FWHM of the Gaussian fitted data is $1.17 \pm 0.03$ mm. This value closely agrees with the coherence length of the two-photon source (~ 1.26 mm), estimated from the pump duration and the group velocity dispersion (GVD).

In particular, the analysis of the two-photon interference fringes obtained from the proposed interferometric measurement using the theoretical description of the joint spectral intensity of the photon pair source (see Methods for details) can yield the asymmetric bi-photon wave functions. Figure 5 shows the bi-photon wave functions and their coincidence count rates for both the frequency separable and entangled cases. We assume that $\tau_1 = \Delta x_1 / c$, $\tau_2 = -\Delta x_2 / c$ and that the profiles of the bi-photon wave functions are Gaussian, while the bi-photon wave functions are filtered by CWDMs in the actual experimental setup shown in Fig. 2. The interference fringes shown in Fig. 5b,d are asymmetric since the center frequencies $\omega_{c,1}$ and $\omega_{c,2}$ of $\Phi(\omega_1,\omega_2)$ are different. This indicates that the period of the interference pattern of $\tau_i$ is $\omega_{c,i}$ when $\tau_j$ is fixed. The interference pattern obtained by varying $\tau = \tau_1 = -\tau_2$ is a HOM type (phase-insensitive) interference, and it exhibits oscillations related to the spatial beating effect represented by $\cos(\Delta\omega\delta t)$ term as written in Eq. (1).

The experimental results in Fig. 4a correspond to the discrete version of those shown in Fig. 5d for specific values of $\tau_1 = \Delta x_1 / c$. The position correlation of the interference fringe between $\Delta x_1$ and $\Delta x_2$ shows that the measured bi-photon wave function has frequency entanglement. Since the bi-photon wave function is filtered by the CWDM (square type), the interference patterns in Fig. 4a are the form of a sinc function. In our experiment, the two-photon coherence length, which is the width of interference pattern for $\Delta x = \Delta x_1 = -\Delta x_2$, is estimated by the visibilities of the patterns shown in Fig. 4b. In the actual experiment, the interference pattern is very dense, and therefore, we consider simple examples in Fig. 5.

In conclusion, we experimentally demonstrated nonlocal two-photon interference between pairs of photons in a fully unfolded and the most general form of the two-photon interferometer including the HOM scheme with non-overlapping two-photon amplitudes and involving spatially separated N00N-like two-photon states, which was carried out by employing two separate but identical photon pair sources. Nonlocal interference between photon pairs was observed even when two photons of a pair did not arrive at the same BS at the same time; this situation cannot be treated using the classical wave theory, and therefore, the situation has no classical analogy. The observed interference fringes simultaneously showed single-photon and two-photon coherence properties in a single interferometer setup. Using the proposed interferometric measurement, we can obtain the joint spectral intensity via the inverse Fourier cosine transform of the full interferogram. Quantum interferences between correlated photons can be utilized to explore quantum communication technologies such as entanglement-based quantum communication, single-photon quantum switching, and single-photon quantum routing by virtue of distributed nonlocal quantum correlations[23-25].

## Methods

**Characterization of photon pair source.** The full characterization of the photon pair source in our experiment can be obtained by simply measuring the single and coincidence counting rates as a function of the applied pump power. For an average pump power of 20 mW, the coincidence efficiency (coincidence to single ratio) was measured to be about 4.7% (the fibre coupling efficiency including optical loss was around 31.3%), and the coincidence to accidental coincidence ratio was about 2.68. The pair production probability per pulse is obtained by dividing the accidental coincidence by the actual coincidence ratio, which was estimated to be 0.37 per pulse in our experiment. For the pulse-pumped SPDC photons accidental coincidences are mainly caused by redundant multiple pair events per pulse.

**HOM interference between two independent photons.** Since the two incoming photons have no time correlations, the HOM effect is rarely observed when the two photons arrive from each source simultaneously. In the experiments, when the two individual photons arrived at FBS1 (FBS2) at the same time, the coincidence counting rate between two detectors was reduced to 0.33 in raw visibility under perfect experimental conditions[11]. From Fig. 3a, the two-fold raw visibility and dip width was measured to be 4.48 ± 0.17% and 0.95 ± 0.05 mm, respectively. The visibility was degraded, and the dip width was broadened due to timing jitter between the independent photon-pair sources[12]. The timing jitter can be caused by the duration time of the pump pulse (3.5 ps) and by the GVD in the SPDC crystals. In our experiment, the timing uncertainty by GVD in a 10 mm-long PPLN crystal was estimated to be about 2.34 ps, which corresponds to 3.42 nm in terms of spectral bandwidth. Unlike the HOM fringe shown in Fig. 3a, the single-photon timing jitter does not affect the interference between photon pairs (Fig. 3b).

**Theoretical description of experimental results and joint spectral intensity.** The quantum state of photon pairs from separate source $\Phi_A$ and $\Phi_B$ (Fig. 1c) has the superposed form of

$$|\Psi\rangle = \iint d\omega_1 d\omega_2 \Phi_A(\omega_1,\omega_2) a^\dagger_{S_A}(\omega_1) a^\dagger_{L_A}(\omega_2)|0\rangle + \iint d\omega_1 d\omega_2 \Phi_B(\omega_1,\omega_2) a^\dagger_{S_B}(\omega_1) a^\dagger_{L_B}(\omega_2)|0\rangle,$$

(4)

where $\Phi(\omega_i,\omega_j)$ denotes the bi-photon wave function, $a^\dagger_i(\omega_i)$ the creation operator of frequency $\omega_i$ at path $i$, and $|0\rangle$ the vacuum state. If $\Phi(\omega_1,\omega_2)$ is separable such as $\Phi(\omega_1,\omega_2) = \Phi_1(\omega_1)\Phi_2(\omega_2)$ then Eq. (4) is the same as the Eq. (2). However, in general, the bi-photon wave function of photon pairs from SPDC is not separable. The two-photon coincidence counting rate between D1 and D3 is proportional to the time-averaged value of the photon detection probability that is defined as

$$P_{D_1 D_3}(t_1,t_3) \propto \left|\langle 0|\hat{E}^{(+)}_{D_3}(t_3)\hat{E}^{(+)}_{D_1}(t_1)|\Psi\rangle\right|^2, \quad (5)$$

where $\hat{E}^{(+)}_{D_i}(t_i)$ denotes the positive part of the electric field operator at $t_i$ in $D_i$ and is superposed form of electric field operators of each input modes ($S_A$, $S_B$) or ($L_A$, $L_B$)[1]. If we assume that each path has its own optical delay lines $\tau_{S_A}$, $\tau_{S_B}$, $\tau_{L_A}$, and $\tau_{L_B}$, then the coincidence counting rate of D1 and D3 is given by

$$G_{13}^{(2)}(\delta\tau_S, \delta\tau_L) = 1 - \text{Re}\left[\Gamma(\delta\tau_S, \delta\tau_L)\right],$$
$$\Gamma(\delta\tau_S, \delta\tau_L) = \iint d\omega_1 d\omega_2 \Phi_A(\omega_1, \omega_2) \Phi_B^*(\omega_1, \omega_2) e^{-i(\omega_1 \delta\tau_S + \omega_2 \delta\tau_L)}, \quad (6)$$

where $\delta\tau_S = \tau_{S_A} - \tau_{S_B}$ and $\delta\tau_L = \tau_{L_A} - \tau_{L_B}$. If the two-photon sources are identical, i.e., $\Phi_A(\omega_1, \omega_2) = \Phi_B(\omega_1, \omega_2)$, $\Gamma(\delta\tau_S, \delta\tau_L)$ denotes the Fourier transform of the joint spectral intensity, i.e., the absolute square of the bi-photon wave function. Figure 5 shows two examples of bi-photon wave functions $\Phi(\omega_i, \omega_j)$ (a: separable, b: non-separable) and their coincidence count rates $G_{13}^{(2)}(\delta\tau_S, \delta\tau_L)$. The experimental results in Fig. 4a correspond to the discrete version of Fig. 5d for specific values of $\delta\tau_s$.

This result is similar to Eq. (6) in Ref. 15, but not identical. The result of Ref. 15 is the Fourier transform of the *symmetrized* bi-photon wave function, which is given as

$\Gamma(\tau_1, \tau_2) = \iint d\omega_1 d\omega_2 \Phi(\omega_1, \omega_2) \Phi^*(\omega_2, \omega_1) e^{-i(\omega_1 \tau_1 + \omega_2 \tau_2)}$, and therefore the scheme in Ref. 15 cannot be applied to asymmetric bi-photon wave functions. The expression for $\Gamma(\delta\tau_S, \delta\tau_L)$ clearly indicates that the two-photon coherence length is not affected by the separation of correlated photons of a pair, but that it is decided by the relative distance between two pairs. In the experiment results, $\delta\tau_S$ and $\delta\tau_L$ are corresponding to $\tau_1 = \Delta x_1 / c$, $\tau_2 = -\Delta x_2 / c$ as mentioned in discussion section. Additionally, we can estimate the joint spectral intensity $|\Phi(\omega_1, \omega_2)|^2$ of the source from the full interferogram of the experimental result, $G_{13}^{(2)}(\delta\tau_S, \delta\tau_L)$. The full interferogram $G_{13}^{(2)}(\delta\tau_S, \delta\tau_L)$ is related to the real part of the Fourier transform of $|\Phi(\omega_1, \omega_2)|^2$, which corresponds to the Fourier cosine transform. Thus, the inverse Fourier cosine transform of $1 - G_{13}^{(2)}(\delta\tau_S, \delta\tau_L)$ gives the joint spectral intensity $\left|\Phi(\omega_1, \omega_2)\right|^2$.

**Figure 1. Generalized two-photon quantum interferometer.** (a) Hong-Ou-Mandel interferometer with a long time delay between two photons. Two photons in a pair incident from different input ports do not arrive at the beamsplitter (BS) simultaneously. However, the total state of two photons in two input ports is in a symmetrically superposed state in space and time. (b) Interferometer involving the N00N-like two-photon state. In this scheme, the two photons do not arrive at the same BS simultaneously. Unfolded two two-photon probability amplitudes can arise from the independent pair sources. (c) Conceptual schematic of our nonlocal quantum interference experiment with two independently generated pairs of photons under Einstein's locality condition. Two correlated photons with angular frequencies $\omega_1$ and $\omega_2$ originate from the separate sources $\Phi_A$ or $\Phi_B$, respectively. In the interferometer, one photon travels along the short path $S_A$ ($S_B$) while the other photon travel along the long path $L_A$ ($L_B$), and subsequently each photon arrives at different BSs (BS1 and BS2) to undergo nonlocal two-photon interference.

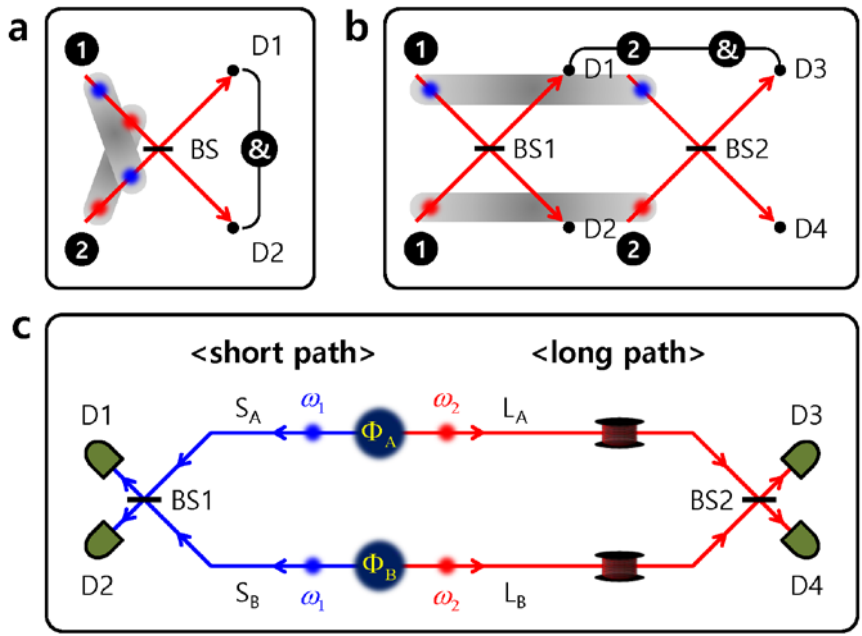

**Figure 2. Experimental setup.** A picosecond pulse laser pumps two type-0 periodically poled lithium niobate (PPLN) crystals, where spontaneous parametric down-conversion (SPDC) occurs under the non-collinear frequency-nondegenerate quasi-phase matching condition. Pump: picosecond mode-locked fibre laser (3.5 ps, 20 MHz, 775 nm, 20 mW); H: Half-wave plate; P: linear polarizer; L1, L2: spherical lenses with focal length of 200 mm; PPLN: periodically-poled lithium niobate crystal (length 10 mm, grating period 19.2 μm, temperature 40°C); DM: dichroic mirror (T1550 nm/R775 nm); L3: aspherical lens with focal length of 8 mm; PC: polarization controller; ODL: optical delay line; FBS: fibre beamsplitter 50/50; CWDM: coarse wavelength division multiplexing filter (bandwidth 18 nm); D1-D4: gated mode single photon detection modules (Id Quantique id-210, id-201, and id 200).

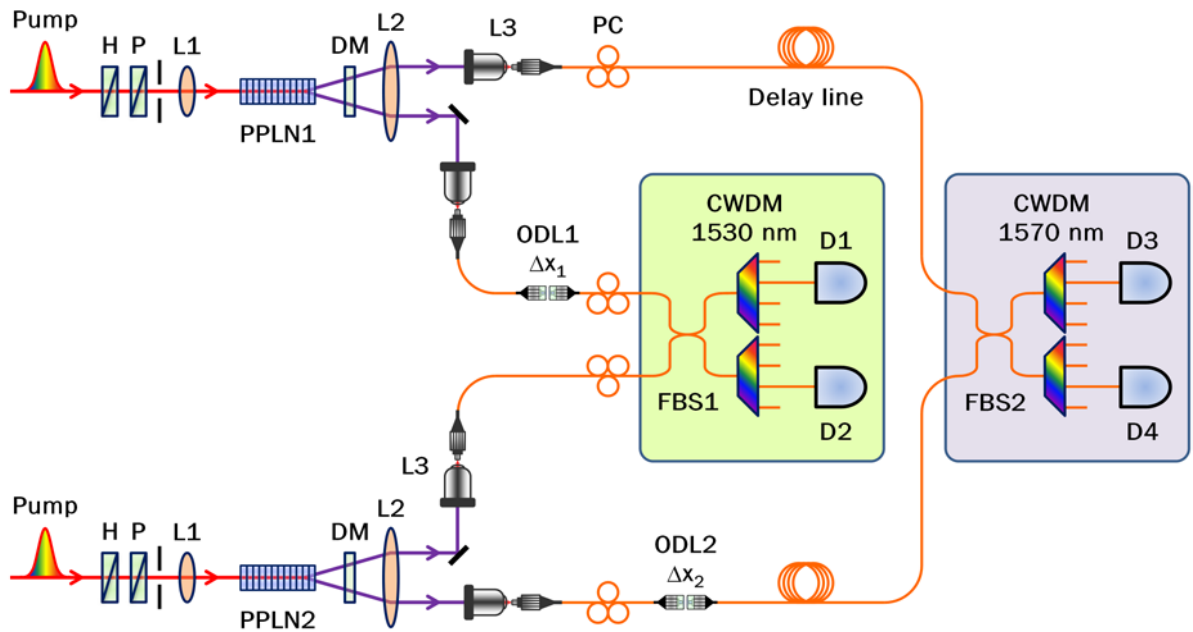

**Figure 3. Two-photon interference fringes.** (a) Result of Hong-Ou-Mandel dip measurement with individual single photons from the two independent sources. Weak two-photon bunching is observed in the coincidence measurements with single photon detectors D1 and D2. (b) Two-photon interference fringe measured in the two-fold coincidences between D1 and D3 as a function of the path-length difference $\Delta x_2$ for the condition of $\Delta x_1 = 0$. (c) Two-photon interference fringes measured at $\Delta x_2 \approx 0$ ($V$=99.57±1.09%). The accidental coincidences are subtracted from the raw data. A high-visibility two-photon interference fringe is observed when the two paths of two individual photons in different pairs are perfectly equalized within the coherence length of the single-photon wave packet.

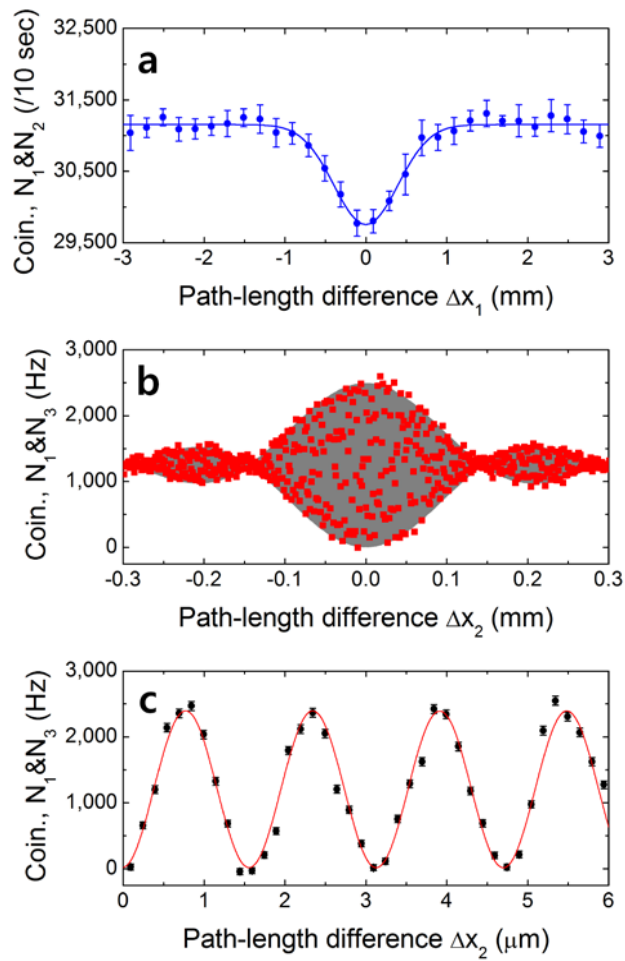

**Figure 4. Two-photon interference fringes and visibility.** (a) Measured two-fold net coincidences between D1 and D3 in which the two-photon interference fringes are obtained for various positions of $\Delta x_1$. (b) Fringe visibility as a function of $\Delta x_1$. The FWHM is estimated to be 1.17 ± 0.03 mm from the Gaussian fitting.

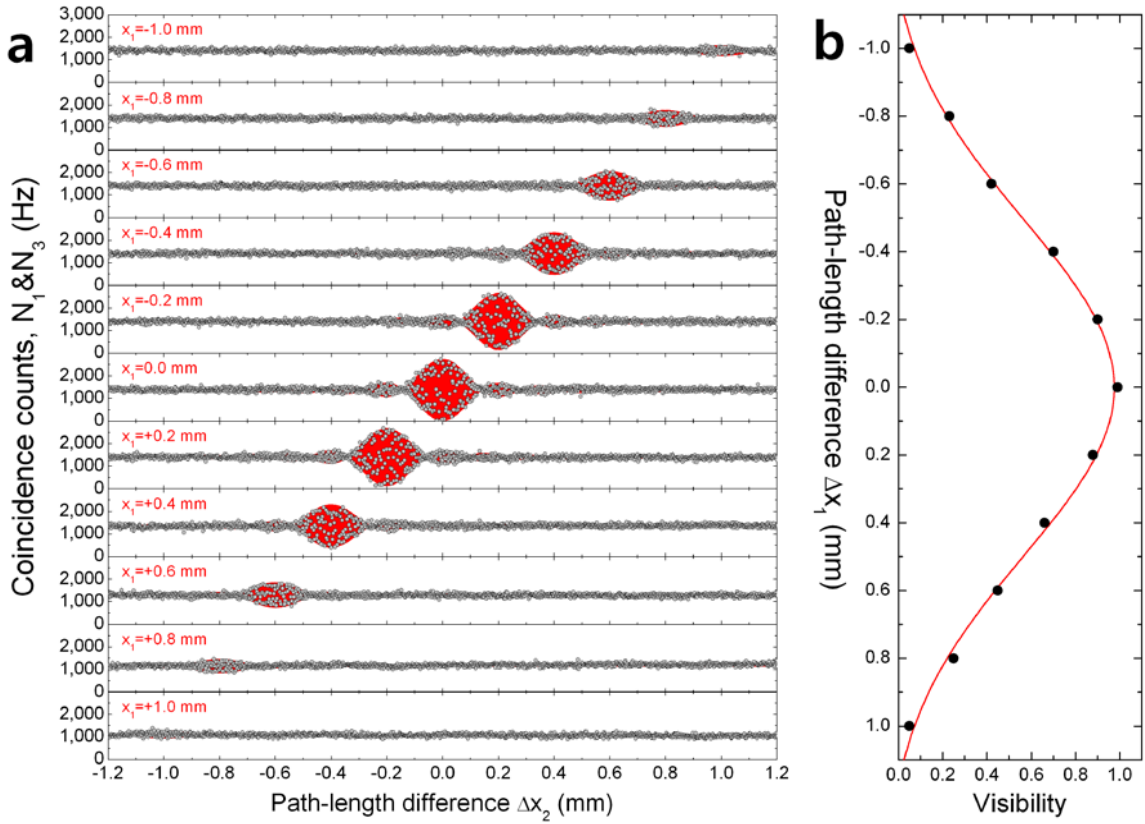

**Figure 5. Asymmetric bi-photon wave function and their coincidence counting rate.** (a) and (b) correspond to the case of the frequency-separable two-photon state, and (c) and (d) correspond to the case of the frequency-entangled two-photon state, where $\tau_1 = \delta\tau_S$ and $\tau_2 = \delta\tau_L$.

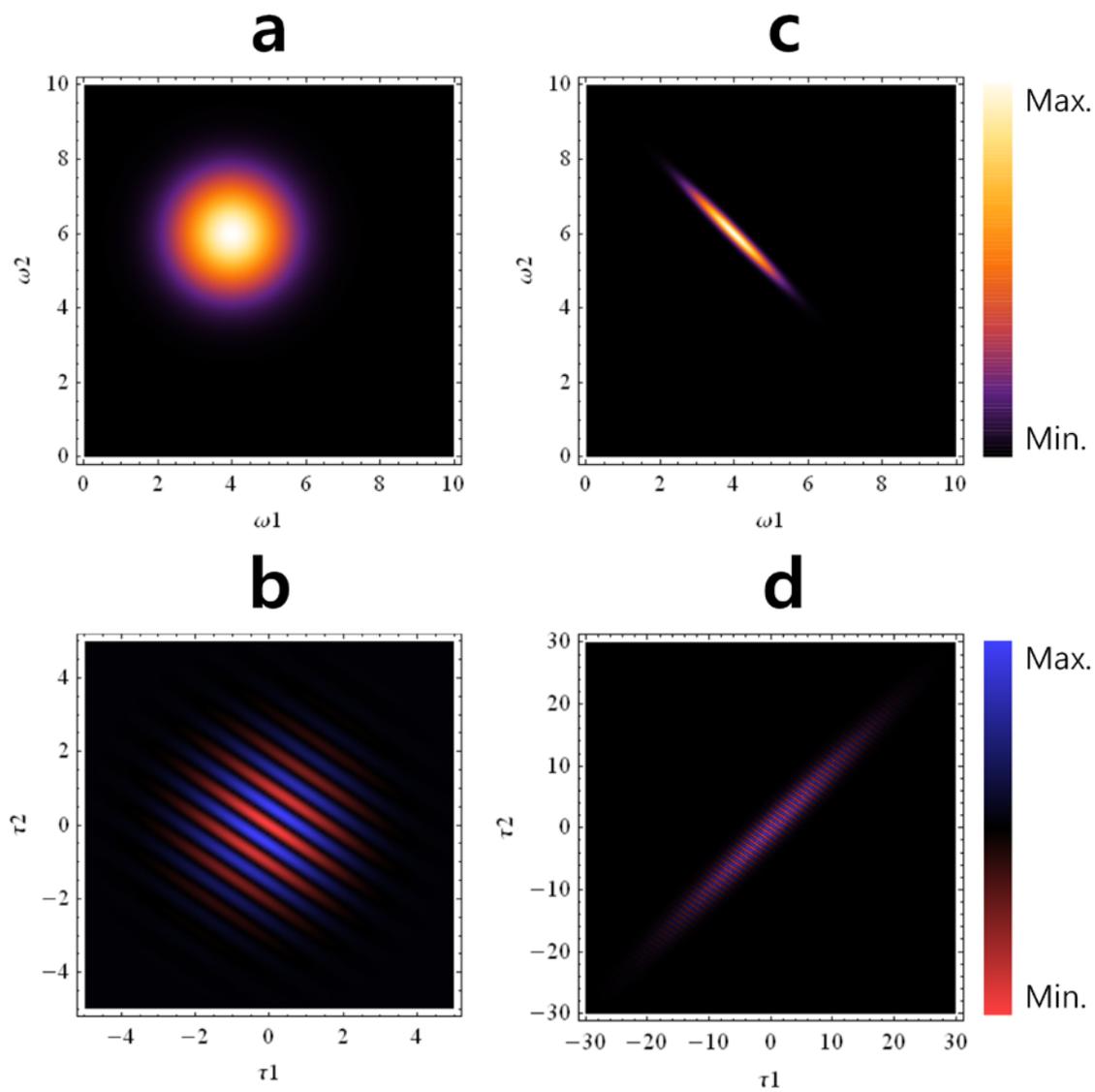